# IMAGING METHOD FOR INTERFACE RHEOLOGICAL CHARACTERIZATION

Nicolas Abi Chebel*/**, Olivier Masbernat*, Frédéric Risso***, Pascal Guiraud****, Florent Ravelet***, Christine Dalmazzone**, Christine Noïk**

* Laboratoire de Génie Chimique, UMR 5503 CNRS-INPT-UPS, 5 rue Paulin Talabot, 31106 Toulouse Cedex1, France
** Institut Français du Pétrole,  Direction Chimie et Physico-Chimie Appliquées, 1 et 4 avenue de Bois Préau, 92852 Rueil-Malmaison, France
*** Institut de Mécanique des Fluides de Toulouse, UMR 5502 CNRS-INPT-UPS, Allée C. Soula, 31400 Toulouse, France
**** UMR 5504, UMR 792 Ingénierie des Systèmes Biologiques et des Procedés, CNRS, INRA, INSA, 135 Avenue de Rangueil, F-31077 Toulouse Cedex 4, France

**KEYWORDS**:
**Main subject(s):** *multi-phase flow*
**Fluid:** *hydrodynamics*
**Visualization method(s):** *shadow method*
**Other keywords:** *oscillation modes, surfactants, crude oil*

**ABSTRACT** : The present work investigates free damped oscillations of an oil drop in water after its release from a capillary tube. Both pure heptane drops and diluted crude oil drops are considered (in the second case the interface is covered by amphiphilic species, natural components of crude oil). Shadowgraph images of the drops are taken by means of a high speed camera and the drop contour is detected by image processing. The axisymmetric drop shape is then decomposed into spherical harmonics, which constitute the eigenmodes of oscillations predicted by the Rayleigh-Lamb theory. Time evolution of each mode is then obtained. The frequency and the damping rate of the principal mode (n=2) are accurately determined and compared with theoretical values for an immobile clean drop oscillating around spherical shape. For pure heptane drops, theoretical value of the frequency agrees well with experiments whereas the damping rate is significantly underestimated by theory. The experimental results clearly show that the different modes are coupled. Energy is thus transfered from mode n=2 to n=3, which probably explains the observed enhancement of the damping rate. The effect of the interface viscoelastic behaviour, induced by adsorbed amphiphilic species on the free oscillations was examined. No significant effect was observed in the experiments conditions (small amplitude oscillations and moderate aging).

## 1 Introduction

During crude oil extraction, water is often co-produced. The water-cut may exceed 80 vol. % for wells at final stages of life. Water-in-oil emulsions are thus produced when liquid-liquid flow encounters cross-section restrictions as the wellhead choke valve. These emulsions are stabilized by amphiphillic species which are natural components of crude oil. Understanding the emulsion formation process is an essential step towards the conception and optimisiation of separation methods.

Drop breakup has been previously studied for a turbulent flow in a vertical column provided with a concentric restriction [1]. Results have shown that the deformation and breakup process depended both on the flow hydrodynamics and the drops own dynamics. The role of the interface response in the breakup mechanism was previously displayed by Risso and Fabre [2] for bubbles in a turbulent field. The Hinze-Kolmogoroff force balance approach [3, 4] was pointed out to be insufficient to predict experimental data (break-up probabilities and highest stable diameter). The drop/bubble deformation was described by a linear oscillator forced by the Lagrangian turbulent Weber number. The drop-oscillator is therefore characterized by an oscillation frequency and a damping rate for each oscillation mode, predicted by the Rayleigh-Lamb theory [5, 6]. An accurate determination of the latter properties is thus necessary for the development of break-up models. The present work investigates free damped oscillations of an oil drop in water after its release from a capillary tube. It aims to study the dynamics of both clean interfaces and interfaces covered with amphiphilc species and to asses the adsorbed species effect on the drops dynamics.

A pre-deformed fluid particle immersed in a stagnant medium performs oscilations driven by periodic exchange between kinetic and potential energies [7]. Oscillation eigenmodes were determined by Rayleigh [5] and Lamb [6]. Their calculations are based on the potential flow theory and assume the oscillations are of low amplitude. Each eigenmode corresponds to a spherical harmonic and is described by two integers n and m. This study focuses on axisymetric oscillation modes, for which $m=0$.

For an axisymetric mode n, the interface local position in spherical coordinates can be expressed using $r(\theta,t)$, referred to as the local radius:

$$r(\theta,t)=a\left[1+A_n\cos(\omega_n t)\cdot P_n(\cos\theta)\right] \tag{1}$$

where $a$ is the non-deformed drop radius, $A_n$ the amplitude of the nth mode oscillations, $\omega_n$ the angular frequency and $P_n(\cos\theta)$ the $n^{th}$ order Legendre polynomial of $\cos\theta$, given by

$$P_n(\cos\theta)=\frac{1}{2^n n!}\cdot\frac{d^n}{d(\cos\theta)^n}\left[(\cos^2\theta-1)^n\right] \tag{2}$$

The oscillation frequency expression for a given eigenmode n is:

$$\omega_n=2\pi\ f_n=\sqrt{\frac{(n-1)n(n+1)(n+2)}{(\rho_d(n+1)+\rho_c n)}\cdot\frac{\sigma}{a^3}} \tag{3}$$

where $\rho_c$ and $\rho_d$ are respectively continuous and dispersed phases densities and $\sigma$ the interfacial tension.

The particle shape can be written as a sum of spherical harmonics and for small amplitude deformation, the shape evolution is given by the superimposition of independant linear harmonics modes, and written as:

$$r(t,\theta)=a\left[1+\sum_n A_n\cos(\omega_n t)\cdot P_n(\cos\theta)\right] \tag{4}$$

In the absence of external forcing, oscillations are damped and the particle tends to recover an equilibrium spherical shape. In the case of low viscosity fluids, the oscillation amplitude decrease as a time exponential:

$$A_n=A_n(0)\cdot\exp(-\beta_n t) \tag{5}$$

The damping rate $\beta_n$ is derived from the dissipation corresponding to the potential-flow fields [6]. For a two fluids system, its expression is [8]

$$\beta_n=\frac{(n+1)(n-1)(2n+1)\mu_d+n(n+2)(2n+1)\mu_c}{\left[\rho_d(n+1)+\rho_c n\right]\cdot a^2} \tag{6}$$

where $\mu_c$ and $\mu_d$ are the viscosities of the continuous and disperse phases respectively.

## 2 Materials and methods

### 2.1 Experimental setup and liquid-liquid systems

Two liquid phase systems were considered: clean interface system and covered interface system. In both cases, the continuous phase consisted of tap water. *n*-heptane (purity 96 %) was used for the first system, while crude oil diluted in *n*-heptane (10 % vol.) was used for the covered interface system. In fact, crude oil contains amphiphilic species (mainly asphaltenes and resins) that adsorb at the liquid-liquid interface, providing it with viscoelastic behaviour.

| Properties at 25°C | Continuous phase (water) | Disperse phase | |
|---|---|---|---|
| | | Heptane | Diluted crude oil |
| Density $\rho$ (kg/m³) | 997 | 685 | 697 |
| Viscosity $\mu$ (Pa.s) | 9·10⁻⁴ | 4.7·10⁻⁴ | 4.7·10⁻⁴ |
| | Water / heptane | Water / diluted crude oil | |
| Interfacial tension $\sigma$ (mN.m⁻¹) | 47 | 25 (interface age 20 min) | |

**Table 1.** Main physical properties of the liquid-liquid system

The experimental setup consists of a plane sided flask filled with continuous phase. Oil drops are released from a vertical U-shape capillary tube immersed in the continuous phase and connected to a 2 mL syringe. Diluted crude oil drops are allowed to age before their release in order to permit the diffusion of the amphiphilic species towards the interface and the formation of an interfacial network [7 bouriat]. Two aging times were considered: 5 min and 20 min. Shadowgraph images of the drops are taken by means of a high speed camera (Photron APX), at a rate of 1000 frames/second and a resolution of 512 x 1024 pixels. Light source consisted of a light-emitting diodes plate. The outer diameter of the capillary tube is used for image calibration.

### 2.2 Drop shape analysis

The drops shape is analyzed by image processing using an algorithm developed under Matlab® environment. Captured grayscale images are stored as matrices of 1024 x 512 elements; each element is assigned with the correspondent pixel greyscale level, ranging between 0 and 256. After background subtraction and image binarization, the drops projection is detected.

The polar coordinates of the boundary pixels are determined and an interface equation is fitted to the detected line. Assuming that the drops are axisymetric, the interface equation in spherical coordinates is established for each image:

$$r = \sum_{i=0}^{5} k_i \cos^i \theta \qquad (7)$$

where $\theta$ is the azimuthal angle. Equation (7) is then rearranged and the local radius written as a linear combination of Legendre polynomials,

$$r = \sum_{i=0}^{5} B_n P_n(\cos\theta) \qquad (8)$$

where $B_n$ the multiplying coefficient corresponding to the nᵗʰ order Legendre polynomial of $\cos\theta$, i.e. the $n^{th}$ order spherical harmonic. Spherical harmonic corresponding to n=0 and n=1 represent, respectively the volume variation and the drop translation. They were not considered in our case which involves incompressible fluids and where the boundaries coordinates are defined with respect to the drop gravity centre. Equation 8 is normalized by the radius $a$, leading to:

$$r(t,\theta)=a\left[1+\sum_n \tilde{A}_n P_n(\cos\theta)\right], \qquad \tilde{A}_n=\frac{B_n}{a} \qquad 2\leq n\leq 5 \qquad (9)$$

For $2\leq n\leq 5$, $A_n$, plotted against time, constitutes the $n^{th}$ eigenmode contribution to the drop free oscillations. The oscillation frequency and the damping rate can then be measured for different eigenmodes.

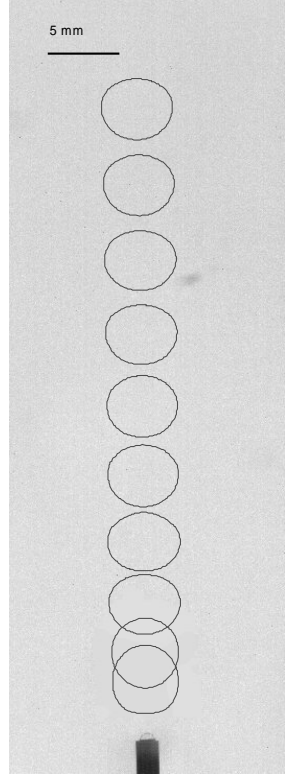

**Fig. 1.** Reconstitution of the drop projection. Time between 2 consecutive plots: 30 ms

## 3 Results and discussion

### 3.1 Clean interface system

Signals corresponding to 4 oscillation modes of a 4.6 mm diameter heptane drop are shown in Fig. 2. We can notice a significant contribution of mode 3, while the second mode remains the dominant one. A signal shift, increasing with time, is observed especially for the second mode: the drop average shape, initially spherical, evolves towards an oblate spheroid, due to the drop ascent in the stagnant phase. Time evolution of the second mode is represented in fig. 3, after being centered by subtracting a moving average from the raw time evolution plots. A damped sinusoidal signal is then obtained and can be modeled by:

$$\tilde{A}_2 = A_2(0)\,e^{-10t}\cos(2\pi\cdot 25\cdot t) \qquad (10)$$

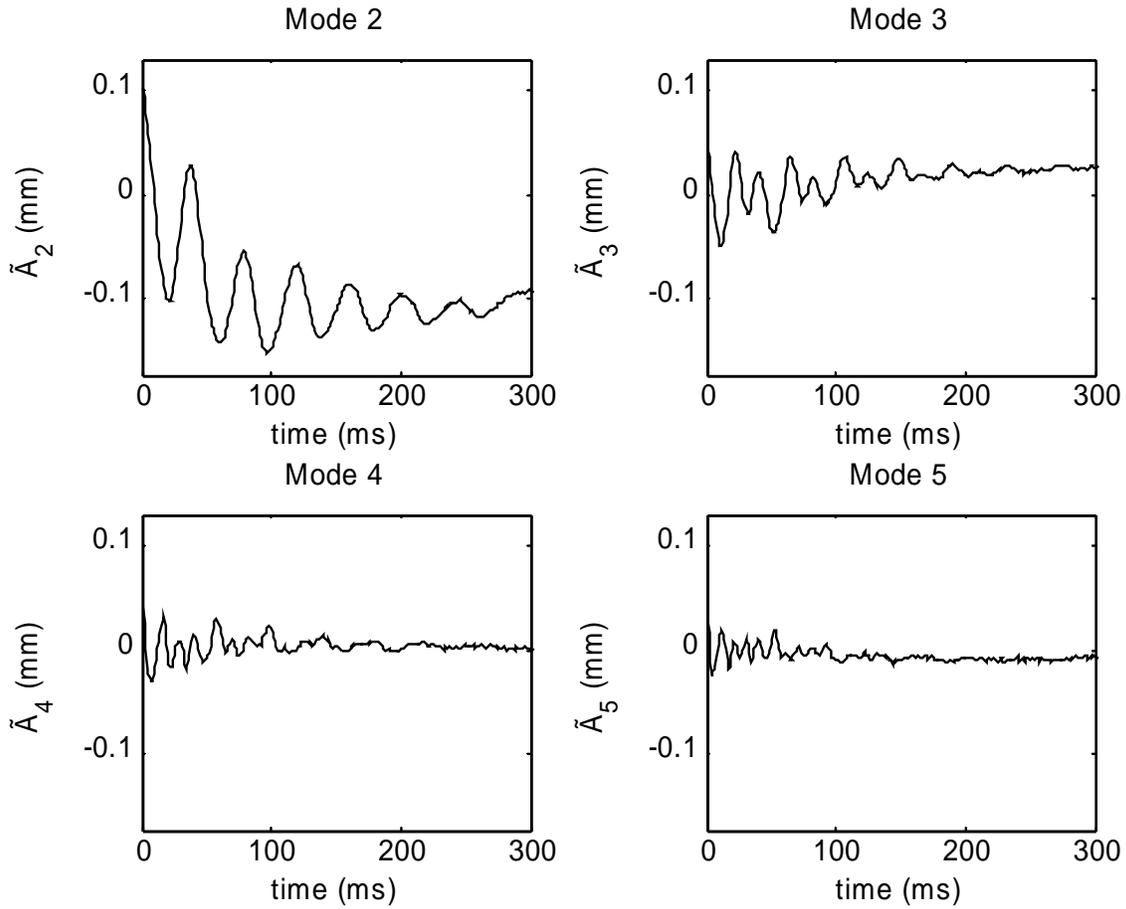

**Fig. 2.** Time evolution of $A_n$ for 4 oscillation modes

The experimental oscillation frequency is 25 Hz and agrees well with the value predicted by the Rayleigh-Lamb theory ($f_2$ = 24 Hz, equation 3). However the mentioned theory underestimates the damping rate (equation 6): the experimental damping rate for the second mode (10 s$^{-1}$) is 5 times larger than the theoretical value ($\beta_2$ = 2s$^{-1}$).

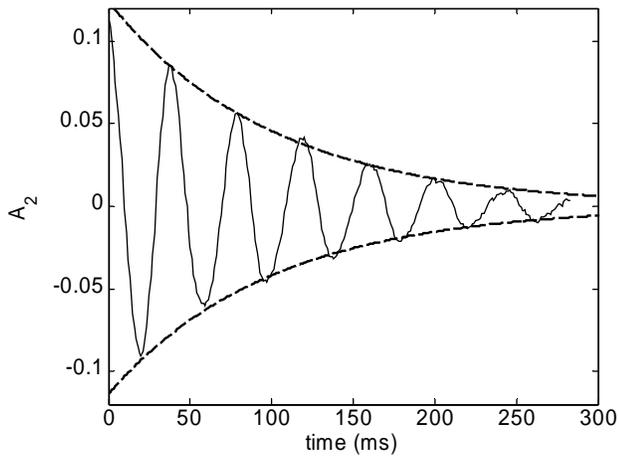

**Fig. 3.** Heptane drop: Centered oscillation signal for mode 2; dashed line: exponential decay (damping rate 10 s$^{-1}$)

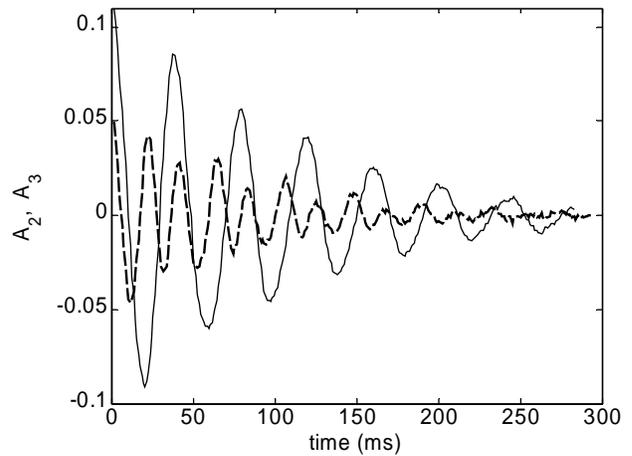

**Fig. 4.** Heptane drop: Centered oscillating signals for mode 2 (continuous line) and mode 3 (dashed line).

Centered oscillation signals for both modes 2 and 3 are represented in figure 4. We notice that mode 3 oscillations are not purely sinusoidal as predicted by the Rayleigh-Lamb theory. However, the highest frequency measured (47 Hz) matches well with the theoretical value (45 Hz).

The signal corresponding to mode 3 is modulated by the second mode signal, which has roughly a twice smaller frequency. Coupling between modes leads to energy transfer from mode 2 to mode 3, which probably explains the observed enhancement of the second mode damping rate.

**3.2 Covered interface system**

Crude oil amphiphilic components adsorb at oil / water interface constituting a viscoelastic network. A variation in the interfacial area induces an interfacial tension variation, which depends on the interface deformation (area variation) and deformation speed. The interfacial dilational viscoelastic behaviour was previously characterized by interfacial rheology experiments involving low frequency forced oscillations (dynamic drop tensiometer [9]). It has been observed that the interface dilational elasticity and viscosity depend on the oscillation frequency and the interface age. The frequency range of the dynamic tensiometer was from 0.1 to 1 Hz, therefore, interfacial elasticity and viscosity values corresponding to the drop natural frequency (tens of Hz) were obtained by extrapolation.

Figure 5 displays the second mode free oscillations for a 20-minute-old diluted crude oil drop. The measured oscillation frequency is 27 Hz, approximately equal to the value predicted by the Rayleigh-Lamb theory, which stands for a uniform and constant interfacial tension. The measured damping rate is 13.76 s$^{-1}$, 3.5 times the theoretical one (3.76 s$^{-1}$), while the ratio between the experimental and theoretical damping rates for the clean interface case was 5. Therefore, no significant effect of the adsorbed material was observed in our experimental conditions, i.e. low amplitude oscillations and moderate aging.

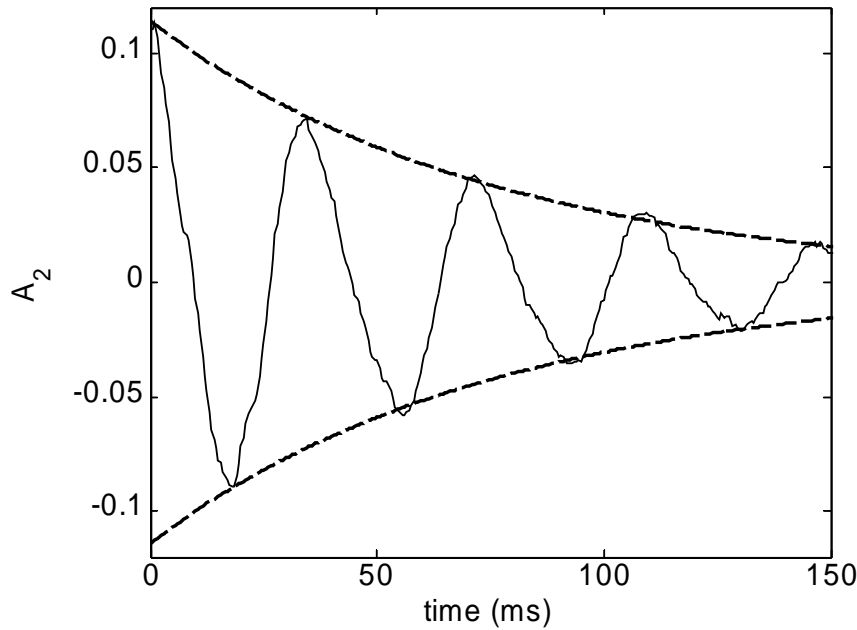

**Fig. 5.** Diluted crude oil drop: Centered oscillation signal for mode 2; dashed line: exponential decay (damping rate 13.76 s$^{-1}$)

In order to explain the obtained results, let us consider the expression of the oscillation frequency (equation 3). The frequency is proportional to the square root of the interfacial tension ($\sigma^{1/2}$), which is assumed constant and uniform all over the interface. When the interface is covered by adsorbed species, the interfacial tension is no longer constant and depends on the interfacial area variation, the interface is then said to have viscoelastic behavior. Rheological constitutive equations establish the relation between the interfacial tension and the interfacial deformation (aria variation) and deformation speed. Assuming the interface behavior is described by the Kelvin-Voigt model, the constitutive equation writes:

$$\sigma = \sigma_0 + \varepsilon \alpha + \eta \dot{\alpha}, \quad \text{with} \quad \dot{\alpha} = \frac{d\alpha}{dt} = \frac{1}{A}\frac{dA}{dt} \quad (11)$$

where $A$ is the interfacial area, $\sigma_0$ the non-deformed drop interfacial tension, $\varepsilon$ the interfacial elasticity and $\eta$ the interfacial viscosity, respectively 30 mN/m and 0.05 mN.s.m$^{-1}$ (values extrapolated to match the drop second mode frequency). Therefore, for small amplitude deformation, the interfacial tension variation is insignificant and the oscillation frequency corresponds to the clean interface case.

On the other hand, the interfacial viscosity contribution in the oscillations damping is evaluated by calculating a surface damping rate, which is proportional to the square of the oscillation amplitude:

$$\beta_n^S = \frac{1}{2(2n+1)} \cdot \frac{(n+1)n}{n\rho_c + (n+1)\rho_d} \cdot \frac{\eta}{a^3} \cdot A_n^2 \quad (12)$$

The surface damping rate (0.03 s$^{-1}$) is also insignificant compared to the volume damping rate (equation 6).

**4 Conclusion**

Drop free oscillations in a liquid at rest have been analysed by video processing. First results have shown the contribution of several eigenmodes in the drop free oscillation, whose frequencies are well predicted by the Rayleigh-Lamb theory. Coupling is observed mainly between modes 2 and 3 and accounts for an energy transfer between the mentioned modes, which could explain that measured damping rates are large compared to those predicted by the above-mentioned theory. The effect of the interface viscoelastic behavior, induced by adsorbed amphiphilic species on the free oscillations was examined. No significant effect was observed for the present experimental conditions, which correspond to oscillations of small amplitude and moderate interface aging.